%% file: main.tex
\documentclass[a4paper,fleqn,usenatbib]{mnras}
\usepackage[T1]{fontenc}
\usepackage{ae,aecompl}
\usepackage{longtable}
\usepackage{makecell}
\usepackage{multirow}
\usepackage{supertabular,booktabs}
\usepackage{upgreek}

\usepackage{graphicx}	
\usepackage{amssymb}	

\usepackage{url}
\usepackage{color}
\usepackage{url}
\usepackage{journals}
\usepackage{array,booktabs,arydshln,xcolor}
\usepackage{colortbl}
\usepackage{xcolor}

\newcommand{\ergs}{erg\,s$^{-1}$}

\newcommand {\be}{\begin {equation}}
\newcommand {\ee}{\end {equation}}

\newcommand{\igr}{IGR~J17591$-$2342}
\newcommand{\sax}{SAX~J1808.4$-$3658}

\newcommand{\tmspa}{PSR~J1023$+$0038}
\newcommand{\tmspb}{PSR~J1824$-$2452I}
\newcommand{\tmspc}{PSR~J1227$-$4853}
\newcommand{\lr}{$L_\mathrm{R}$}
\newcommand{\lx}{$L_\mathrm{X}$}
\newcommand{\thh}{$^{\mathrm{th}}$}
\usepackage{soul}

\title[Radio and X-ray monitoring of \igr]
      {Radio and X-ray monitoring of the accreting millisecond X-ray pulsar \igr\ in outburst}
 \author[Gusinskaia et al.]
    {N.V. Gusinskaia,$^{1,2}$\thanks{E-mail: N.Gusinskaia@uva.nl}
    T.D. Russell,$^{1}$
    J.W.T. Hessels,$^{1,2}$
    S. Bogdanov,$^{3}$
    N. Degenaar,$^{1}$
\newauthor A.T. Deller,$^{4}$
    J. van den Eijnden,$^{1}$
    A.D. Jaodand,$^{5}$
    J.C.A. Miller-Jones$^{6}$
\newauthor  and R. Wijnands$^{1}$\\
$^1$Anton Pannekoek Institute for Astronomy, University of Amsterdam, Science Park 904, 1098 XH Amsterdam, The Netherlands\\
$^2$ASTRON, the Netherlands Institute for Radio Astronomy, Postbus 2, 7990 AA, Dwingeloo, The Netherlands\\
$^3$Columbia Astrophysics Laboratory, Columbia University, 550 West 120th Street, New York, NY, 10027, USA\\
$^4$Centre for Astrophysics and Supercomputing, Swinburne University of Technology, P.O. Box 218, Hawthorn, VIC 3122, Australia\\
 $^5$Cahill Center for Astrophysics, 1216 E. California Blvd, California Institute of Technology, Pasadena, CA 91125, USA\\
 $^6$International Centre for Radio Astronomy Research, Curtin University. GPO Box U1987, Perth, WA 6845, Australia}

\date{Accepted xxxx xxx xx.  Received xxx xxxx xx; in original form 2019}

\pubyear{2019}

\begin{document}
\label{firstpage}
\pagerange{\pageref{firstpage}--\pageref{lastpage}}
\maketitle

\begin{abstract}

\igr\ is a new accreting millisecond X-ray pulsar (AMXP) that was recently discovered in outburst in 2018. Early observations revealed that the source's radio emission is brighter than that of any other known neutron star low-mass X-ray binary (NS-LMXB) at comparable X-ray luminosity, and assuming its likely $\gtrsim 6$\,kpc distance.  It is comparably radio bright to black hole LMXBs at similar X-ray luminosities. In this work, we present the results of our extensive radio and X-ray monitoring campaign of the 2018 outburst of \igr. In total we collected 10 quasi-simultaneous radio (VLA, ATCA) and X-ray ({\it Swift}-XRT) observations, which make \igr\ one of the best-sampled NS-LMXBs.  We use these to fit a power-law correlation index $\beta = 0.37^{+0.42}_{-0.40}$ between observed radio and X-ray luminosities (\lr\ $\propto$ \lx$^{\beta}$).  However, our monitoring revealed a large scatter in \igr's radio luminosity (at a similar X-ray luminosity, \lx\ $\sim 10^{36}$\,\ergs, and spectral state), with \lr\ $\sim 4 \times 10^{29}$\,\ergs\ during the first three reported observations, and up to a factor of 4 lower \lr\ during later radio observations. Nonetheless, the average radio luminosity of \igr\ is still one of the highest among NS-LMXBs, and we discuss possible reasons for the wide range of radio luminosities observed in such systems during outburst.  We found no evidence for radio pulsations from \igr\ in our Green Bank Telescope observations performed shortly after the source returned to quiescence.  Nonetheless, we cannot rule out that \igr\ becomes a radio millisecond pulsar during quiescence.  

\end{abstract}


\section{Introduction}
\label{sec:intro}

Low-mass X-ray binaries \citep[LMXBs;][]{Lewin1997} are systems in which a black hole (BH) or neutron star (NS) primary accretes material from a low-mass companion star that overflows its Roche lobe and forms of an accretion disk around the primary.  Quasi-simultaneous radio (\lr) and X-ray (\lx) luminosity measurements of such systems can help probe the interplay between accretion inflow: X-rays, which probe the inner-most parts of the accretion disk and matter transfer onto the primary in the case of NSs; and outflow: flat\footnote{Spectral index $\alpha \sim 0$, where the flux density $S_{\nu}$ at frequency $\nu$ scales as $S_{\nu} \propto \nu^{\alpha}$.} or inverted spectrum radio emission from a partially self-absorbed, collimated jet\footnote{Although not all observed radio emission is consistent with coming from a compact jet \citep{Bogdanov2018}.} \citep[e.g.,][]{MIGFEN2006,Gallo2012,Corbel2013}.    

In the last few years, new and upgraded radio telescopes --- like the Karl G. Jansky Very Large Array (VLA), the Australia Telescope Compact Array (ATCA), and MeerKAT --- coupled with flexible X-ray observatories such as the Neil Gehrels Swift X-ray Telescope ({\it Swift}-XRT), have provided the sensitivity and rapid response required to probe a steadily increasing number of known Galactic NS-LMXBs. This includes accreting millisecond X-ray pulsars \citep[AMXPs;][]{Tudor2017}, where the accretion flow is channeled by the NS's magnetic field (producing coherent X-ray pulsations: e.g., \citealt{Wijnands1998}; see \citealt{Patruno2012} for a review), as well as non-pulsing NS-LMXBs --- both of which have been observed in the hard and soft X-ray spectral states \citep[e.g.,][]{Gusinskaia2017}.  

NS-LMXBs have been observed during their bright (X-ray luminosity \lx\ $\sim 10^{36-38}$\,\ergs) accretion outbursts \citep[e.g.,][]{MIGFEN2006}, and as they fade back to quiescence. This has allowed the \lr---\lx\ relationship in the hard X-ray state to be studied for both individual systems \citep[e.g.,][]{MIG2003,TUD2009,Tetarenko2016} as well as statistically for the population as a whole \citep[e.g.,][]{Gallo2018}.  Previous studies have modelled the radio--X-ray luminosity scaling in the simplistic form \lr\ $\propto$ \lx$^{\beta}$.  Recent work by \citet{Gallo2018} finds $\beta = 0.44^{+0.05}_{-0.04}$ and $\beta = 0.59\pm0.02$ for the NS- and BH-LMXB populations, respectively.  \citet{Gallo2018} also found that NS-LMXBs are, on average, $\sim20$ times less radio luminous compared to BH-LMXBs at the same \lx.

Overall, radio observations during the hard X-ray state are consistent with the presence of a compact, partially self-absorbed jet during outburst \citep{BZ1977,BP1982}.  However, both NS- and BH-LMXBs show a wide range of \lr\ for comparable \lx, as well as roughly a two-order-of-magnitude spread in average \lr\ between systems.  Furthermore, individual systems show different power-law slopes compared to the global averages \citep{Tudor2017,Gusinskaia2019}.  There is currently no clear physical picture explaining this observed diversity, but perhaps it is unsurprising given that NS-LMXBs have a range of NS spin rate, magnetic field, and magnetospheric inclination with respect to the disk.  In the case of AMXPs, \citet{Tudor2017} question whether it is even appropriate to fit a single \lr---\lx\ correlation over a broad luminosity range ($> 1$\,dex, i.e. an order of magnitude) because these systems show very large scatter.  The same could be said for NS-LMXBs in general, where to date there is no demonstrated tight power-law correlation for individual systems over a large ($> 2$\,dex) luminosity range \citep{Gusinskaia2019}.  

AMXPs like \sax\ directly demonstrated that NSs with millisecond spin periods exist in accreting systems \citep{Wijnands1998}, thereby providing strong observational support for the recycling scenario in which radio millisecond pulsars (MSPs) are formed via the transfer of mass and angular momentum in such systems \citep{Alpar1982,Radha1982}.
This scenario has been further demonstrated by the relatively recently identified class of transitional millisecond pulsars (tMSPs), which switch on multi-year timescales between radio pulsar and accretion-disk states \citep[e.g.][]{archibald2009,Papitto2013,Stappers2014}.  These have been observed using high-sensitivity multi-wavelength campaigns during their prolonged low-luminosity disk states (\lx\ $\sim 10^{32-34}$\,\ergs; one of the tMSPs, M28I, has shown a full AMXP-like outburst). While their bright (compared to other NS-LMXBs) and rapidly variable radio emission was initially ascribed to a compact, partially self-absorbed synchrotron jet \citep{DEL2015}, the discovery of an anti-correlation between radio and X-ray luminosity in strictly simultaneous observations suggests a different physical mechanism \citep{Bogdanov2018}. This is supported by recent multi-wavelength observations, which suggest rapid ejection due to a pulsar wind (\citealt{Papitto2019}, Jaodand et al 2019 in prep.). 

Despite the great increase in NS-LMXB \lr---\lx\ measurements in recent times, additional observations --- both of new systems, as well as higher cadence and/or strictly simultaneous radio---X-ray observations of previously observed systems --- are needed to identify possible correlations between radio brightness and system parameters like the orbital inclination, neutron star mass, spin rate, magnetic field strength, and orbital period.  For example, \citet{Eijnden2018} detect a faint radio jet from a highly magnetised ($B > 10^{12}$\,G), slowly spinning NS accreting at a rate above the theoretical Eddington limit.  They argue that the magnetic field strength of the NS could have a significant role in dictating the radio brightness, but whether this is also applicable to lower magnetic field systems is unknown.  At the same time, recent numerical simulations are investigating the role of the NS magnetic field \citep{Parfrey2016,Parfrey_magnsph_2017,Parfrey_GR_2017}.

\subsection{\igr}

On August 10\thh---11\thh, 2018, during its continuing survey of the Galactic bulge and centre, the International Gamma-Ray Astrophysics Laboratory ({\it INTEGRAL}) X-ray and gamma-ray telescope discovered a new transient, \igr\ \citep{ATel_igr_discovery}.  The high absorption seen in the X-ray spectrum \citep{ATel_igr_swift_original,ATel_igr_chandra} suggested that this source is likely roughly at the distance of the Galactic centre ($\sim$8 kpc).  \citet{Nowak2019} further investigate and discuss the source distance, and conclude that it is very likely at the distance of the Galactic bulge or beyond ($> 6$\,kpc).  \citet{ATel_igr_optical} could not identify an optical counterpart, but \citet{ATel_igr_NIR} detected an associated transient near-infrared source.

Initial radio observations using ATCA measured a high flux density, typical of that seen in BH-LMXBs when the X-ray luminosity is $\sim 10^{36}$\,\ergs\ \citep{Russell2018,ATel_igr_radio}.  However, timing  analysis of follow-up Nuclear Spectroscopic Telescope Array ({\it NuSTAR}) and Neutron Star Interior Composition Explorer ({\it NICER}) observations revealed 527-Hz coherent X-ray pulsations, showing that \igr\ was an AMXP in  outburst \citep{ATel_igr_pulsations,Sanna2018}. Other properties of the source were also similar to those typically seen for other AMXPs (and NS-LMXBs) in outburst, e.g., a relatively low peak X-ray luminosity (\lx\ $\sim 10^{36}$\,\ergs) and hard spectrum (photon index $\Gamma \sim 2$).

\begin{figure*}
\centering
\includegraphics[width=\textwidth]{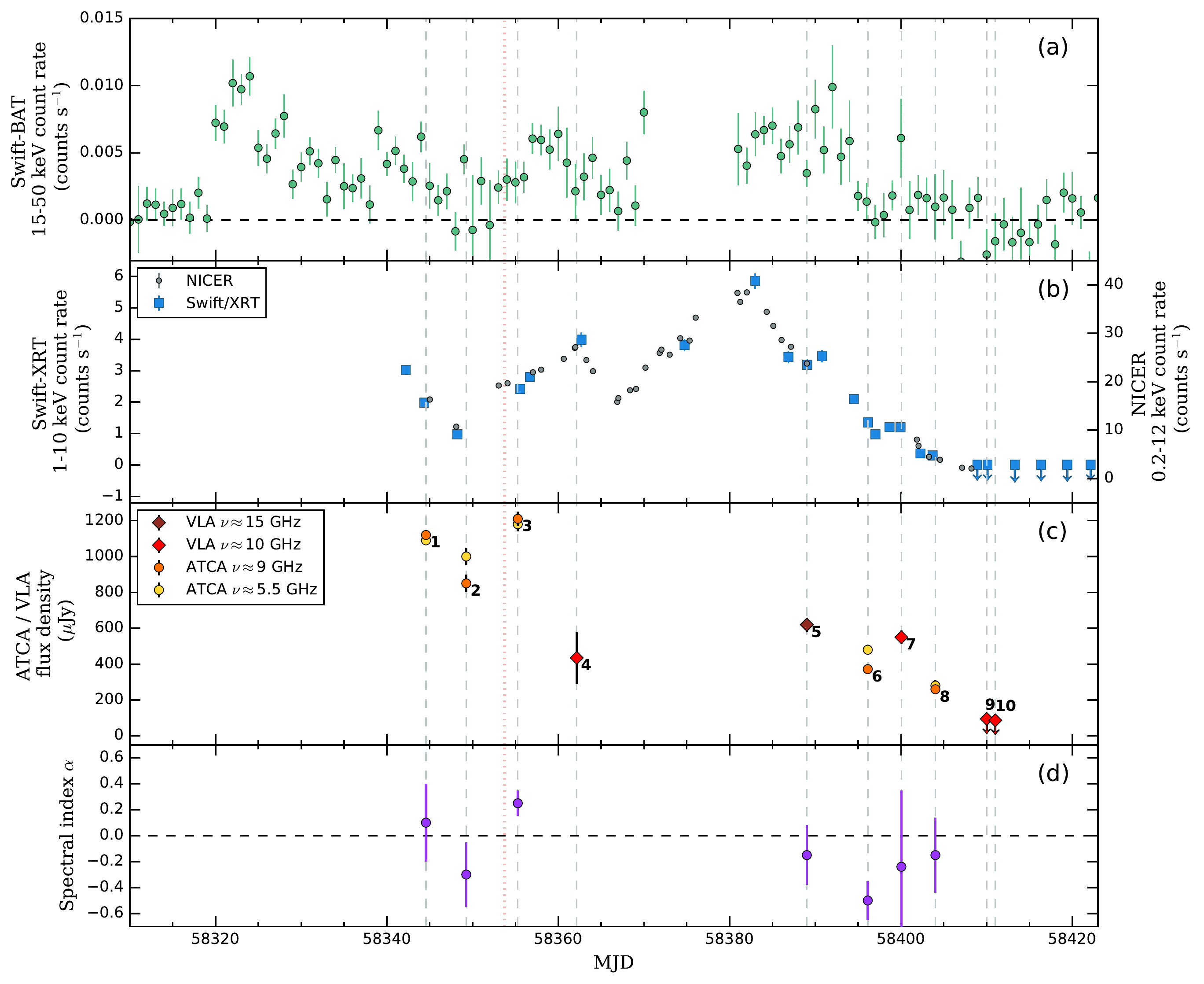}

\caption{X-ray and radio light-curves during the 2018 outburst of \igr.  Downward pointing arrows indicate 3$\sigma$ upper limits in all cases. \textbf{Panel (a):} {\it Swift}-BAT ($15-50$\,keV) daily X-ray light curve. \textbf{Panel (b):} Blue squares represent the {\it Swift}-XRT ($1-10$\,keV) X-ray light-curve (using the left-hand axis). Grey circles represent the {\it NICER} ($0.2-12$\,keV) X-ray light-curve (using the right-hand axis).  \textbf{Panel (c):} Radio light-curve. Circles and diamonds represent ATCA and VLA observations, respectively; their colour shade represents the frequency. \textbf{Panel (d):} Inferred radio spectral index $\alpha$, where $S_{\nu} \propto \nu^{\alpha}$. Vertical grey dashed lines indicate the epochs of radio observations. The vertical red dotted line indicates the {\it Chandra} observation epoch in which a wind was detected \citep{Nowak2019}.}\label{fig:igr_lcs}
\end{figure*}

The outburst displayed a light-curve with multiple brightness peaks over the course of its approximately 3-month duration (Figure~\ref{fig:igr_lcs}).
Based on archival data from the {\it Swift} Burst Alert Telescope ({\it Swift}-BAT), \citet{ATel_igr_bat_original} reported that the outburst had actually started on roughly July 22$^{\rm nd}$ (about 3 weeks prior to the discovery by {\it INTEGRAL}), peaked on July 25\thh\ and decayed thereafter.  As such, the initial radio observations of \citet{Russell2018} were performed close in time, but weeks after the first bright peak of the outburst (Figure~\ref{fig:igr_lcs}), and did not sample the later X-ray bright peaks that were observed \citep{ATel_igr_integral_rebr,ATel_igr_deep_rebr,ATel_nicer_update}.

\citet{Nowak2019} presented high-resolution spectroscopy using the {\it Chandra X-ray Observatory}, performed during the start of the second peak of the outburst (see red dotted vertical line in Figure~\ref{fig:igr_lcs}) and demonstrated evidence for an outflowing wind in \igr.  The AMXP \sax\ also showed evidence for a wind \citep{Pinto2014}.  Such X-ray winds are usually observed during the soft X-ray states of LMXBs, in which the jet is typically quenched \citep{Miller2006,Neilsen2009,Ponti2012}.  However, \citet{Homan2016} found that for high-luminosity NSs (Z-sources), high-velocity winds are observed during the hardest spectral state, where radio jet emission is the brightest.

X-ray timing observations provided a precise orbital ephemeris and showed that \igr\ has an orbital period $P_b = 8.8$\,hr and a minimum companion mass 0.42\,M$_\odot$ \citep{Sanna2018}.  This, along with its relatively high spin rate, makes \igr\ quite similar to the well-established tMSP systems \tmspa\ \citep{archibald2009}, \tmspb\ \citep{Papitto2013}, and \tmspc\ \citep{Bassa2014} --- which are all eclipsing radio millisecond pulsars with relatively high-mass ($> 0.2$\,M$_\odot$) companions \citep[such systems are termed `redbacks'][]{Roberts2013}.  Overall, \igr\ is a possible tMSP candidate (although the observation of more tMSP-like phenomenology would strengthen the case), and might become a rotation-powered radio pulsar during quiescence.

In this paper, we present quasi-simultaneous radio---X-ray observations that span \igr's 2018 outburst from shortly after its discovery by {\it INTEGRAL} until it faded into quiescence.  This is the largest-available radio-monitoring data set; it includes and significantly extends the work of \citet{Russell2018}.  We use this to constrain the \lr---\lx\ correlation for \igr\ and to quantitatively compare its average radio luminosity and scatter, in the $4 \times 10^{35}$\,\ergs\ $<$ \lx\ $< 10^{37}$\,\ergs\ range, with those of AMXPs, non-pulsing NS-LMXBs, and BH-LMXBs.  We also present high-time-resolution radio observations acquired using the Robert C. Byrd Green Bank Telescope (GBT), with which we search for a radio pulsar signal in the weeks following the outburst --- in order to investigate whether \igr\ switches on as a radio pulsar in quiescence. 

\section{Observations and data analysis}
\label{sec:obs}
\input{table1.tex}

\subsection{Radio continuum data and analysis}

\subsubsection{VLA}

\igr\ was observed with the VLA at 5 epochs (project ID: 18A-374) on September $1^{\rm st}$, 28\thh\ and October 9\thh, 19\thh\ and 20\thh, 2018 (see Table~\ref{tab:igr_obs}).  All observations were taken at X-band ($8-12$\,GHz), except for the second observation (September 28\thh), which was recorded using the Ku-band receiver ($12-17$\,GHz). Each observing session had a duration of $\sim 1$\,h including calibration scans, providing $\sim 30$\,min of on source time. During all 5 epochs the array was in D configuration (synthesized beam $\sim 12$\arcsec).   We used 3C~286 as the flux and bandpass calibrator, and we used J1820$-$2528 and J1744$-$3116 as the phase calibrator in the X-band and Ku-band observations, respectively. All observations were processed using the Common Astronomy Software Application (CASA\footnote{\url{https://casa.nrao.edu/}} version 5.4.1; \citealt{2007ASPC..376..127M}). During our first three observations the VLA was still recovering from an extended power outage, and these data had unusual artifacts, which precluded using the standard EVLA-pipeline calibration routine. Instead, all initial calibration and flagging was done manually, following standard calibration procedures within CASA. In order to mitigate effects from diffuse emission in the field we excluded short baselines ($< 12 {\rm k}\lambda$) from the analysis and performed imaging using the Briggs weighing scheme\footnote{\url{http://www.aoc.nrao.edu/dissertations/dbriggs/}} with robust parameter set to 0.

\subsubsection{ATCA}

We also observed \igr\ using the Australia Telescope Compact Array (ATCA) on October 5\thh\ and 13\thh, 2018 (see Table~\ref{tab:igr_obs}; the typical observing durations were $\sim 4$\,hr), under the project code CX413. The observations were taken at central frequencies of 5.5 and 9\,GHz, with 2\,GHz of bandwidth at each frequency band, and the array was in 6A configuration. We used PKS~1934$-$638 for flux and bandpass calibration, while J1752$-$225 was used for phase calibration. Data were flagged, calibrated and imaged following standard procedures within CASA. Imaging was carried out using a Briggs robust parameter of 0 to reduce the effects of diffuse emission in the field. In this work, we also use the three ATCA observations of \igr\ presented by \citet{Russell2018}.

For all ATCA and VLA radio epochs we extracted the target flux density by fitting for a point source in the image plane using the {\tt imfit} task within CASA using the source position reported in \citet{Russell2018}.  These radio flux densities, along with spectral indices determined from a least-squares fit to multi-frequency measurements, are reported in Table~\ref{tab:igr_obs} and Figure~\ref{fig:igr_lcs}.  

We attempted time-resolved radio analysis to search for intra-observation variability during our VLA and ATCA observations. Our observations constrain any brightness variations to within 10\% of the average flux density.  However, we note that --- due to the combination of poor instantaneous uv-coverage, a complex field, and low source brightness --- our sensitivity to source brightness changes on short timescales is poor. In particular, for ATCA observations we were unable to search for variability on timescales shorter than $\sim 30$\,minutes (and on some days only down to 1\,hour intervals). At the same time, any differences in the source's flux density inferred from 10 or 15-minute scans in our VLA observations were dominated by the relatively large uncertainties on the flux density of the source caused by the diffuse emission in the field. Therefore, we cannot constrain possible variability on few-minute timescales (or less), which has been observed in some NS-LMXBs \citep{Bogdanov2018}.

\subsection{X-ray data and analysis}

In order to obtain the full context of the X-ray luminosity and spectrum during \igr's outburst, we produced light-curves using {\it Swift}-XRT, {\it Swift}-BAT and {\it NICER} observations (Figure~\ref{fig:igr_lcs}). The {\it Swift}-XRT $1-10$\,keV light-curve was produced using the online {\it Swift}-XRT data products generator (\citealt{xrt_prod_lc1,xrt_prod_lc2}; target IDs: 10803, 10804, and 10899). The {\it Swift}-BAT daily light-curve ($15-50$\,keV) was obtained from the {\it Swift}-BAT transient monitor results provided by the {\it Swift}-BAT team \citep{Krimm2013}. Lastly, the {\it NICER} $0.2-12$\,keV light-curve was obtained from the online available event files (observational ID: 12003101). Using the {\tt NICERDAS} package in {\tt HEASOFT} (v. 6.24\footnote{\url{https://heasarc.gsfc.nasa.gov/docs/software/heasoft/}}) we filtered out `hot' Focal Plane Modules (FPMs 34, 14 and 54) as well as times within Sun, Moon or South Atlantic Anomaly constraints. We then extracted the light-curve using {\tt XSELECT}.

\subsubsection{{\it Swift}-XRT spectral analysis}

To give quasi-simultaneous X-ray flux and spectral context to our radio measurements, we selected 10 {\it Swift}-XRT observations (target IDs 10804 and 10899) that were performed within 0.6 days of the VLA and ATCA observations (see Table~\ref{tab:igr_obs} and Figure~\ref{fig:igr_lcs}). All such observations were carried out in the Photon Counting (PC) mode. Their observation IDs, epochs and inferred spectral parameters are listed in Table~\ref{tab:igr_obs}.

For this X-ray spectral analysis we used {\tt HEASOFT} (v. 6.24). First, we ran the {\tt XRT$\_$PIPELINE} on each observation for basic data reduction and calibration. Then, we extracted the spectrum for all 10 epochs using {\tt XSELECT}. The source spectrum was extracted from a 30-pixel-radius circular area, centred on the source position (taken from \citealt{Russell2018}), and the background spectrum was extracted from an annular region with 30-pixel inner radius and 60-pixel outer radius, also centred at the source position. Additionally, following \citet{pile-up}, we excluded the $3-5$ central pixels of the source region in order to account for pile-up in the case of observations with count-rate higher than one count per second. To correct for this --- and also for known artefacts on the CCD as well as the response of the telescope --- we produced the ancillary arf-file using {\tt XRTMKARF} together with exposure maps for each event and the response file 'swxpc0to12s6$\_$20130101v014.rmf'.

{\tt XSpec} (version 12.10.0c; \citealt{1996ASPC..101...17A}) was used to perform the spectral model fit and extract the source flux. We used photons in the $0.4-10$\,keV range and a simple absorbed power law model ({\tt TBabs * powerlaw} in {\tt XSpec}) for all observations. We fixed the hydrogen column density parameter in the {\tt TBabs} model to $N_{\rm H} = (4.4\pm 0.2) \times 10^{22}$\,cm$^{-2}$ (letting it vary within the errors), which was provided by the extensive analysis of \citet{Nowak2019}. We used the additional {\tt cflux} convolution model to determine the unabsorbed $1-10$\,keV fluxes and their associated errors for each epoch. The inferred fluxes, photon indices as well as $\chi^2_{\nu}$ values and degrees of freedom (dof) for each observation are listed in Table~\ref{tab:igr_obs}.

\subsection{GBT searches for radio pulsations}

Subsequent to \igr's accretion outburst, which ended roughly on October 20\thh, 2018, we performed three GBT observations (project code: GBT18B-353), which were spaced weekly on October 28\thh, and November 4\thh\, and 11\thh, 2018 (see Table~\ref{tab:igr_gbt_obs}).  Each observing session was $\sim 1$\,hr in duration with $\sim45$ minutes on source, which is about 8\% of the orbital period ($P_{b} = 8.8$\,hr; \citealt{Sanna2018}). It is important to note that `redback' radio pulsars often exhibit eclipses from intra-binary material blown off of the companion star by the pulsar wind \citep[e.g.][]{archibald2009}. In order to avoid such eclipses, we performed our GBT observations at the times when \igr\ was close to inferior conjunction, using the ephemeris of \citet{Sanna2018} to predict orbital phase (Table~\ref{tab:igr_gbt_obs} lists the orbital phase range of each observation). All observations were done in a high-time-resolution search mode. We used the GBT S-band ($1.73-2.60$\,GHz) receiver and the GUPPI recording backend in its online coherent dedispersion mode.  We recorded full Stokes data over a 700-MHz (epochs 1 and 2) or 800-MHz (epoch 3) band, with 10.24\,$\mu$s samples and 448 (epochs 1 and 2) or 512 (epoch 3) frequency channels of 1.56\,MHz each.  We used a coherent (intra-channel) dispersion measure (DM) of 250\,pc\,cm$^{-3}$ to compensate for the likely high DM of the source.  As we discuss below, there are several pulsars associated with the Galactic bulge that have DM~$177-422$\,pc\,cm$^{-3}$, and our coherent dedispersion trial value is in this range.  The NE2001 electron density model prediction for this line-of-sight is 250\,pc\,cm$^{-3}$ at a distance of 4\,kpc \citep{NE2001}. At a larger distance of 8\,kpc, NE2001 predicts a DM of 635\,pc\,cm$^{-3}$ and a large scattering time at 2\,GHz: $2.5$\,ms, i.e. comparable to the pulse period.  For comparison, the YMW16 electron-density model predicts higher DM: 423\,pc\,cm$^{-3}$ for 4\,kpc and 1088\,pc\,cm$^{-3}$ for 8\,kpc. Note, however, that the scattering timescale along any given line-of-sight is highly uncertain.  Using the DM versus $N_{\rm H}$ relationship of \citet{DMnH2013}, we find that the expected DM is in the broad range of $200-2000$\,pc\,cm$^{-3}$.  Ultimately, the detectability of radio pulsations from \igr\ is challenging given the likely large distance, and also depends on the scattering timescale being favourably low compared to model predictions.

\begin{table}
\caption[All observations]{GBT observations.}

\begin{minipage}{\textwidth}
\renewcommand{\arraystretch}{1.2}
{
\begin{tabular}{@{\extracolsep{-3pt}}l|cccc@{}}
\hline\hline
Date & \#  & MJD & Duration & Orbital phase$^{a}$ \\
     &     &     & (s)  & \\
\hline
2018-10-28 & 1 & 58419.878 & 2402 & $0.71-0.79$\\
2018-11-04 & 2 & 58426.895 & 2759 & $0.85-0.93$\\
2018-11-11 & 3 & 58433.886 & 2653 & $0.91-0.99$\\
\hline\hline
\end{tabular}
}
\begin{flushleft}{
$^{a}$ Inferior conjunction is at orbital phase 0.75.
  }\end{flushleft}
\end{minipage}
\label{tab:igr_gbt_obs}
\end{table}

\begin{figure*}
\centering 
\includegraphics[width=\textwidth]{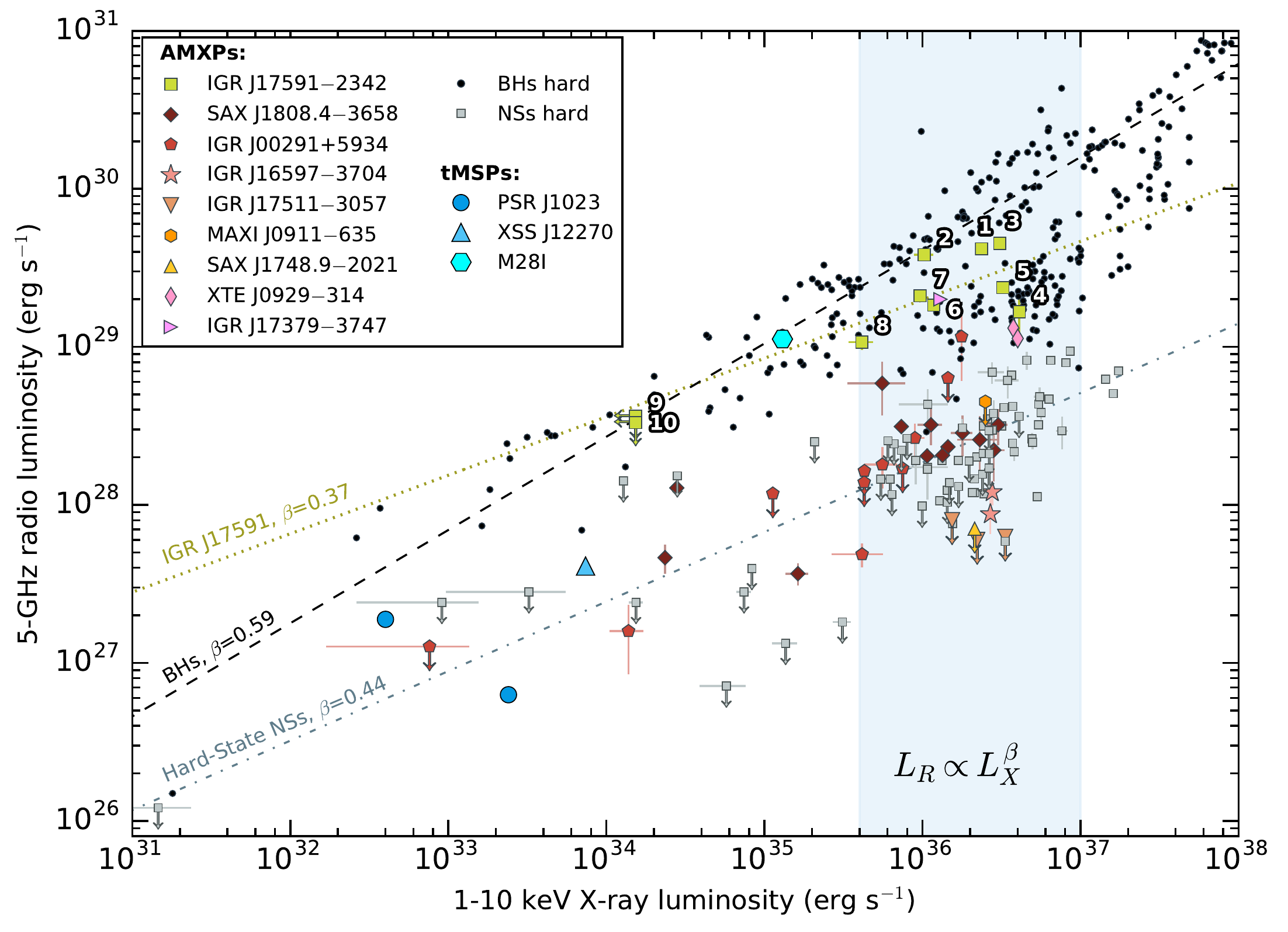} 

\caption{X-ray ($1-10$\,keV) luminosity versus radio (5\,GHz)
  luminosity for BH- and NS-LMXBs. Black circles represent BH-LMXBs; grey squares represent non-pulsing NS-LMXBs; a variety of symbol types and colours are used to represent individual AMXP and confirmed tMSP systems. Radio---X-ray measurements of \igr\ are shown using yellow-green squares (where the assumed distance is 8\,kpc; see \citealt{Russell2018} Figure~2, which plots the measurements for the first 3 epochs for a range of distances) and each observation (Table~\ref{tab:igr_obs}) is labelled with its corresponding ID number.  Data points are taken from \citet{Bahramian2018} for BHs hard; \citet{MIGFEN2006,Tetarenko2016,Gusinskaia2017,Gusinskaia2019} for NSs hard; \citet{Papitto2013} and updated X-ray value from \citet{CotiZelati2019} for M28I, where strictly simultaneous data was used and a time-resolved X-ray spectral analysis was performed; \citet{Hill2011} for XSS~J12270$-$4859; \citet{Bogdanov2018} for PSR~J1023+0038; \citet{Tudor2017} for SAX~J1808.4$-$3658, IGR~J00291$-$5934, and IGR~J17511$-$3057; \citet{Tetarenko2018} for IGR~J16597$-$3704, \citet{Tudor2016ATel} for MAXI~J0911$-$635; \citet{MJ2010ATel}, \citet{Tetarenko2017ATel} for SAX~J1748.9$-$2021; \citet{Migliari2011} for XTE~J0929$-$314; \citet{J17379_ATel1148} for IGR~J17379$-$3747; \citet{Russell2018} and this work for IGR~J17591$-$2342. Correlation tracks for hard state BHs (dashed black line) and NSs (dash-dotted grey line) are defined in \citet{Gallo2018}. The yellow-green dotted line represents the result of our \lr\ $\propto$ \lx$^{\beta}$ fit for \igr. The blue shaded area represents the \lx\ $= 4 \times 10^{35} - 1 \times 10^{37}$\,\ergs\ X-ray luminosity range that we used for our comparison analysis (Figure~\ref{fig:igr_cdf}).}\label{fig:igr_lrlx}

\end{figure*}

We used the {\tt PRESTO} software suite \citep{Ransom2001,Ransom2002,Ransom2003} to search the GBT data for radio pulsations. 
First, we masked narrow-band and impulsive radio frequency interference (RFI) using {\tt rfifind}.  After flagging, and accounting for bandpass, the effective bandwidth of the observations is $\sim 550$\,MHz.  Since the DM of \igr\ is unknown, we must search over this parameter in order to accurately correct for dispersive delay across the observing band.  We created downsampled Stokes I filterbank data with a time resolution of 81.92\,$\mu$s, using {\tt psrfits\_subband}.  We then produced barycentric dedispersed timeseries in a range from $0-1003.2$\,pc\,cm$^{-3}$ using {\tt prepsubband}, according to a dedispersion step plan calculated with {\tt DDplan.py}.  We used DM steps of 0.2, 0.3, 0.5, and 1\,pc\,cm$^{-3}$ and additional downsampling factors of 1, 2, 4, and 8 for DM ranges $0.0-446.4$, $446.4-547.2$, $547.2-787.2$ and $787.2-1003.2$\,pc\,cm$^{-3}$, respectively, producing a total of 3264 time series for each observation.

For each of these time series the search was performed in two ways: 1) a blind Fourier-based periodicity search and 2) direct pulse-phase folding using the known ephemeris \citep{Sanna2018}. For the blind search we used {\tt accelsearch}, which takes into account the Doppler shift of the pulsar period due to its orbital motion.  We used z = 600, to account for a linear drift in the fundamental of up to 600 Fourier bins. For the direct pulse-phase folding search we accounted for potential stochastic orbital variability \citep[typical of redback systems, e.g.][]{Archibald2013} with respect to the reference ephemeris by searching over a $|\Delta$T$_{\rm asc}| < 5$\,s in steps of 0.1\,s.  This resulted in 100 folded profiles per DM.  Both search strategies and the data integrity were verified using the same approach on a 1-minute test scan of PSR~J1802$-$2124, which was easily detected.  The same process failed to identify a signal plausibly associated with \igr.

\section{Results} 
\label{sec:results}

\subsection{X-ray}

During its first detected outburst, \igr\ displayed a complex X-ray light-curve with multiple brightness peaks resolved by {\it Swift}-XRT and {\it NICER} (Figure~\ref{fig:igr_lcs}). Furthermore, {\it Swift}-BAT observations show a bright peak prior to the {\it INTEGRAL} discovery of the source on MJD~58340 \citep{ATel_igr_discovery}.  Using {\it NICER} data \citet{ATel_nicer_update} demonstrated that \igr\ was consistently in a hard state, and that it showed coherent X-ray pulsations \citep{Sanna2018} throughout the outburst.  Our measured X-ray spectra from {\it Swift}-XRT also indicate that \igr\ was in a hard state for the entire span of the radio observing campaign.

Ideally we would also calculate an average daily hardness-intensity diagram (HID, where hardness is usually defined as the ratio of hard $\gtrsim 10$\,keV and soft $\lesssim 10$\,keV fluxes and intensity as the sum of these fluxes) during the outburst, in order to provide further context to the radio measurements.  However, the source faintness and crowded field preclude sufficiently precise measurements using all-sky monitors, and offer no additional useful information.  Furthermore, the high absorption and relatively low source brightness also limits the utility of an HID based on {\it Swift}-XRT or {\it NICER} $\lesssim 10$\,keV data.

The X-ray flux varied by an order of magnitude during the outburst, which translates into a luminosity range \lx\ $= 4 \times 10^{35-36}$\,\ergs\ for an assumed distance of 8\,kpc (\lx\ $= 4 \pi D^2 S$, where $S$ is the observed flux and $D$ is distance to the source). We did not detect the source in {\it Swift}-XRT observations after MJD~58403. For the last observation (target ID: 00010804024) we derived a $3\sigma$ upper-limit on the count rate of $<0.02$\,cnt\,s$^{-1}$ using the online {\it Swift}-XRT data products generator. We converted this count rate upper-limit into a $1-10$\,keV unabsorbed flux upper-limit of $<0.02 \times 10^{-10}$\,\ergs cm$^{-2}$ using {\tt WebPIMMS}\footnote{\url{https://heasarc.gsfc.nasa.gov/cgi-bin/Tools/w3pimms/w3pimms.pl}} with $N_{\rm H} = 4.4 \times 10^{22}$\,cm$^{-2}$ and assumed photon index $\Gamma=2$.

\subsection{Radio continuum}

\igr\ was detected at all 5 epochs of ATCA observations and at 3 VLA epochs (see Table~\ref{tab:igr_obs}); it was not detected during the last two VLA epochs. The measured radio flux densities, 3$\sigma$ upper limits, and spectral indices are listed in Table~\ref{tab:igr_obs}. All fluxes were converted to a 5-GHz radio luminosity, assuming a flat spectral index ($\alpha = 0$, which is consistent with the average observed value) and distance of 8\,kpc, using \lr$= 4 \pi \nu D^{2} S_\nu $, where $\nu$ is the central frequency and $S_\nu$ is the observed flux density.  We found that \lr\ varies throughout the outburst, by a factor of $\sim 5$, and drops significantly (by at least factor of 3) at the end of the outburst. The source was brightest during the first three observations previously presented by \citet{Russell2018} (see Figures~\ref{fig:igr_lcs} and \ref{fig:igr_lrlx}) with \lr\ $\sim 4 \times 10^{29}$\,\ergs, and it decays after MJD~58355, where its radio luminosities were sometimes significantly lower at comparable X-ray luminosity (see Figure~\ref{fig:igr_lrlx}: observations 1,3,4, and 5 at \lx\ $= 3-4 \times 10^{36}$\,\ergs, as well as observations 2,6 and 7 at \lx\ $= 0.9-1.5 \times 10^{36}$\,\ergs).  Given the typically large errors on the spectral indices, we do not detect any clear evolution in the radio spectrum.

\subsection{Radio---X-ray relation}

In order to quantify the differences between \igr\ and other LMXBs, we performed a power-law fit to the radio and X-ray luminosities for different classes of LMXBs, as well as individual sources when the number of available measurements is sufficient ($> 8$ \lr---\lx\ points). Using the {\tt Python} implementation of the \citet{linmix} linear regression algorithm {\tt LINMIX\_ERR}\footnote{\url{https://github.com/jmeyers314/linmix}} we perform a linear fit in logarithmic space, in the form used by \citet{Gallo2014,Gallo2018}:

\begin{equation}
\lg\,L_\mathrm{R}-\lg\,L_\mathrm{R,c} = \lg\,\xi + \beta \left(\lg\,L_\mathrm{X}-\lg\,L_\mathrm{X,c}\right)
\label{lm_eq}
\end{equation}

{\tt LINMIX\_ERR} returns fit results for three free parameters: the power-law index $\beta$, the scaling factor $\xi$ (which, in our case, denotes scaling of the radio luminosity value of the line intercept with respect to a reference radio luminosity $L_\mathrm{R, c}$=5$\times 10^{28}$\,\ergs, for a fixed X-ray luminosity $L_\mathrm{X, c}$=2$\times 10^{36}$\,\ergs) and an additional parameter, $\sigma_{0}$, that accounts for intrinsic random (Gaussian) scatter of the luminosities around the best-fit power law.

In order to directly compare \lr---\lx\ data of \igr\ with other sources, we performed the fit for only data points within the $4 \times 10^{35}$\,\ergs\ $<$ \lx\ $< 10^{37}$\,\ergs\ range. We performed separate fits for all BH-LMXBs, non-pulsing NS-LMXBs, AMXPs (both including and excluding \igr), as well as separate fits for the best-sampled NS-LMXBs: \igr, \sax, and Aql~X-1. The provenance of these data is provided in the caption of Figure~\ref{fig:igr_lrlx}.  The inferred mean radio luminosities $\xi+L_\mathrm{R, c}$ and intrinsic scatters $\sigma_{0}$ of each sample are presented in Figure~\ref{fig:igr_cdf}.  For \igr\ we find a best-fit power-law slope $\beta = 0.37^{+0.42}_{-0.40}$ (Figure~\ref{fig:igr_lrlx}).

\subsection{Radio pulsations search}

In our GBT observations, we found no evidence for radio pulsations from \igr\ during quiescence.  Using the observational parameters, along with the modified radiometer equation \citep{Dewey1985}, we calculate an upper limit on the period-averaged flux density of radio pulsations.  We assume a telescope gain $G = 2.0$\,K/Jy, system temperature $T_{\rm sys} = 30$\,K, integration time $T = 2500$\,s, effective bandwidth $B = 550$\,MHz, pulse duty cycle 10\%, and a minimal S/N = 8 for a plausible detection (this is set to be relatively low because the pulse frequency is known).  The resulting flux density limit is 26\,$\mu$Jy.  This does not account, however, for intra-channel dispersive smearing at DMs far from 250\,pc\,cm$^{-3}$, nor does it account for possible temporal broadening from scattering.

\section{Discussion}
\label{sec:disc}

\begin{figure*}
\centering 
\includegraphics[width=\textwidth]{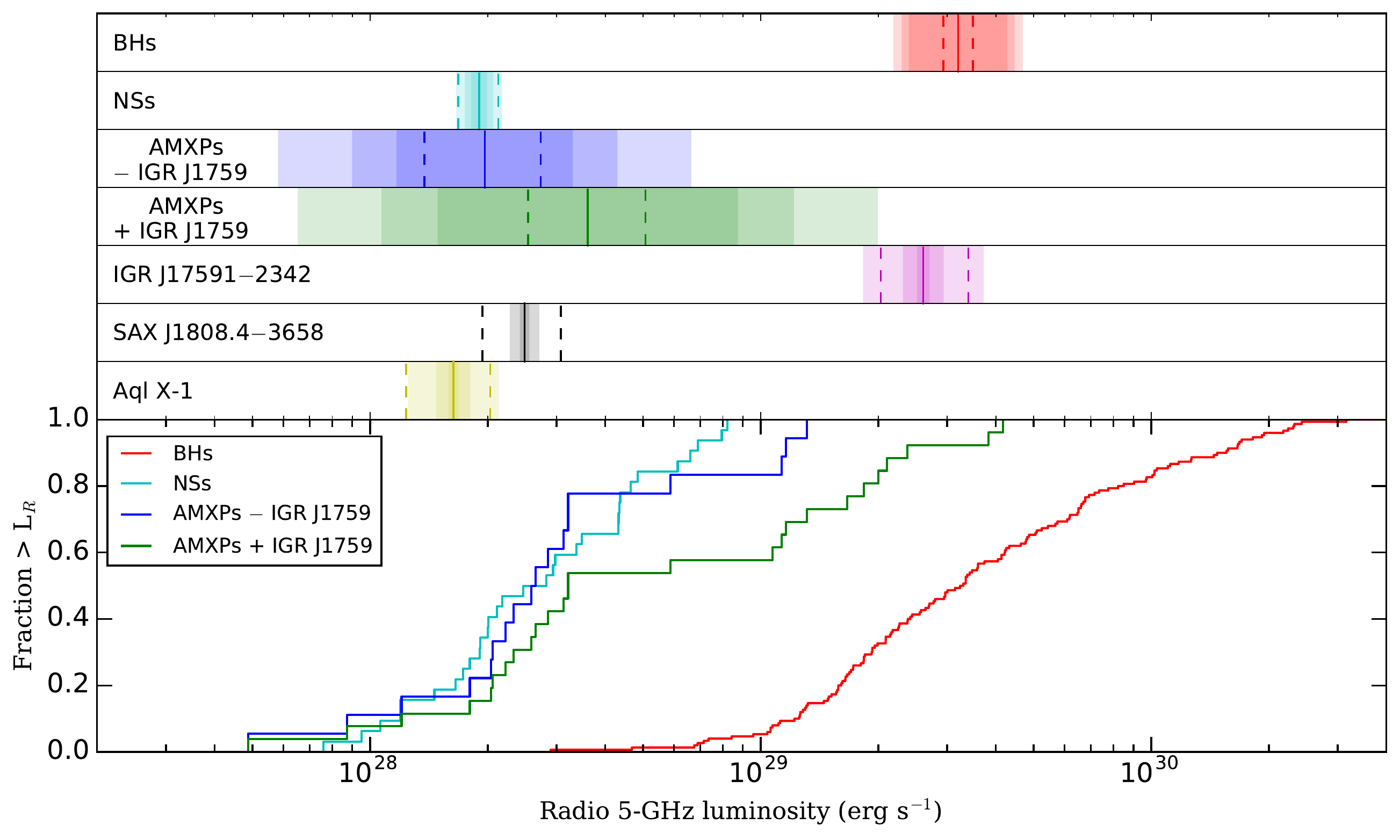} 

\caption{\textbf{\textit{Top:}} Inferred parameters of the {\tt LINMIX\_ERR} fit (Equation~\ref{lm_eq}) to X-ray and radio luminosities of different classes of NS- and BH-LMXBs, as well as individual sources (using an X-ray luminosity range $4 \times 10^{35}$\,\ergs\ $<$ \lx\ $< 10^{37}$\,\ergs). The solid and dashed lines represent values of the radio luminosity intercept $\xi+L_\mathrm{R, c}$ and its uncertainties, respectively. The shaded areas represent the intrinsic scatter $\sigma_0$ of radio luminosities around the best-fit function of the slope, along with its uncertainties. \textbf{\textit{Bottom:}} Cumulative distribution function for 5-GHz radio luminosities of different classes of NS- and BH-LMXBs.}\label{fig:igr_cdf}

\end{figure*}

Next to \sax\ and Aql~X-1, \igr\ is the best-sampled NS-LMXB in the \lx\ $= 4 \times 10^{35} - 1 \times 10^{37}$\,\ergs\ luminosity range that is typically seen during the hard state of NS-LMXB accretion outbursts \citep[see][for a discussion of these sources]{Tudor2017,Gusinskaia2019}.  \igr\ is an AMXP \citep{Sanna2018}, with spin rate, orbital period and companion mass similar to those of, e.g., AMXPs Swift~J1749.4$-$2807 and SAX~J1748.9$-$2021, as well as the established tMSPs \tmspa, \tmspb\ and \tmspc\ \citep[see, e.g., Table~1 of][]{Campana_DiSalvo2018}.

The multi-peaked X-ray light curve of the outburst is not unprecedented for an LMXB --- see, e.g, \citet{Marino2019} for a compendium of AMXP outburst references.  The persistence of coherent X-ray pulsations \citep{Sanna2018} throughout the outburst \citep{ATel_nicer_update} is seen in most AMXPs \citep{Patruno2012} with a few exceptions (Aql~X-1; \citealt{CAS2008}, HETE~J1900$-$2455; \citealt{Patruno2017}, and SAX~J1748.9$-$2021; \citealt{Altamirano2008}).  The fact that \igr\ remained in the hard state, and did not transition to the soft state during the outburst is also commonly observed in AMXPs \citep{Patruno2012,Campana_DiSalvo2018}.  In fact, \igr's most exceptional quality, at least for the purposes of this discussion, is its high radio luminosity during outburst.

Given the large distance to \igr, the lack of detected radio pulsations subsequent to the outburst is arguably unsurprising, and does not rule out an active radio pulsar during quiescence.  There is no evidence that tMSPs are any fainter (or brighter) compared the rest of the known radio millisecond pulsar population.  Radio (millisecond) pulsars in directions a few degrees from the Galactic centre --- and thus in the same general direction and potentially at a comparable distance to \igr\ --- are known \citep[see][who associate these with a roughly Galactic centre distance at $\sim 8$\,kpc based on their observed spin-down rates, which are dominated by acceleration in the Galactic potential]{Perera2019}.  If \igr\ was an equally bright radio pulsar compared with this sample, then we would have easily detected it in our GBT observations.  Radio millisecond pulsars are generally believed to have wide beams \citep{pulsarbook}, so it is unlikely that the emission cone misses Earth. However, we cannot rule out that a large eclipse duration or temporal broadening from scattering in the interstellar medium are responsible for the non-detection.  Most AMXPs fail to show detectable radio pulsations during quiescence \citep[e.g.,][]{Iacolina2010,Patruno2017}, but this does not rule out that they become rotation-powered pulsars \citep[e.g.,][]{Burderi2003}.  Their rotation-powered activity during quiescence will be better probed by the future Square Kilometre Array \citep{Keane2015}, which provides a major boost in sensitivity.  Higher sensitivity allows both shorter observations, which are better suited to detecting pulsations in short-orbital-period system, as well as the opportunity to search at higher radio frequencies where eclipsing is less prominent.

\subsection{Radio---X-ray relation}

While it is unclear whether NS-LMXBs show a single power-law relation between radio and X-ray luminosities, we nonetheless fit for such a correlation so that \igr\ can be better compared with previously studied systems.  

We performed the power-law fit of our \lr--\lx\ measurements of \igr\ restricting to the X-ray luminosity range $4 \times 10^{35}$\,\ergs\ $<$ \lx\ $< 10^{37}$\,\ergs\ (Figure~\ref{fig:igr_lrlx}) because: 1) this range corresponds to that commonly seen in the hard state of NS-LMXBs in outburst and contains the highest density of archival measurements for comparison; and 2) it lowers the chance of mixing different accretion physics (e.g. where a pulsar wind or propellering might have a strong effect), likely occurring at lower X-ray luminosities \citep[e.g.,][]{Tudor2017,Bogdanov2018,Gusinskaia2019}.  Nonetheless, extending the X-ray luminosity range to lower \lx\ would not have a large effect on the measured slopes presented here.

For \igr\ we find a best-fit power-law slope $\beta = 0.37^{+0.42}_{-0.40}$ within the X-ray luminosity range $4 \times 10^{35}$\,\ergs\ $<$ \lx\ $< 10^{37}$\,\ergs\ (which does not include upper-limits, i.e. epochs 9 and 10; see Figure~\ref{fig:igr_lrlx}).  This is consistent with the global fit to hard-state NS-LMXBs found by \citet{Gallo2018}, but the uncertainty on the fit is large because of the significant scatter in the measurements: $\sigma_0 = 0.05^{+0.1}_{-0.04}$\,dex.  Sampling \igr\ using strictly simultaneous radio---X-ray observations would have better constrained or ruled out any \lr---\lx\ correlation.

Given its likely $> 6$\,kpc distance, \igr\ is the radio-brightest AMXP detected thus far.  It is consistent within the scatter with the population of BH-LMXBs (Figures~\ref{fig:igr_lrlx} and \ref{fig:igr_cdf}).  This underscores the importance of using more than just radio brightness in order to classify new X-ray transients as NSs or BHs.

Alternatively, the distance to \igr\ would have to be surprisingly lower than modelled \citep[$< 6$\,kpc;][]{Nowak2019} to bring it more in line with the measured radio luminosities of other NS-LMXBs.  However, this would introduce new questions, because the large $N_{\rm H}$ would have to be ascribed to a local origin \citep[which was deemed unlikely by][]{Nowak2019}, and the outburst would be atypically X-ray faint. 

\sax\ is the only other non-intermittent AMXP with comparable \lr---\lx\ sampling; we find that \igr\ is a factor of 13 times more radio luminous in the $4 \times 10^{35}$\,\ergs\ $<$ \lx\ $< 10^{37}$\,\ergs\ range (Figure~\ref{fig:igr_cdf}).  It is also 16 times more luminous compared to the normally non-pulsing\footnote{Aql~X-1 showed coherent X-ray pulsations for roughly 120\,s out of $\sim 10^6$\,s of observations \citep{CAS2008,nopuls2015}.} Aql~X-1.   

AMXPs show a very large spread in radio luminosities in the considered X-ray region ($4 \times 10^{35}$\,\ergs\ $<$ \lx\ $< 10^{37}$\,\ergs): among them are the faintest and the brightest NS-LMXBs. Almost as luminous as \igr\ are IGR~J17379$-$3747, XTE~J0929$-$314, and the AMXP/tMSP M28I (also known as IGR~J18245$-$2452).  These systems have very few available \lr---\lx\ measurements, however, and are thus interesting targets to explore whether they can be even more luminous. Similarly, \igr\ was first observed only 3 weeks after the start of its 2018 outburst.  The pre-discovery hard X-ray light-curve (Figure~\ref{fig:igr_lcs}, panel~a) suggests that the source may have been the brightest in X-rays during the initial peak, and thus we could have missed its brightest radio emission as well.

Overall, we see that the inclusion of \igr\ in a fit to all AMXPs demonstrates that these span 1.5\,dex in \lr, which is a broader range compared to that observed thus far for non-pulsing NS-LMXBs.  However, this could be a bias due to such systems being preferentially targeted for study.  Without the inclusion of \igr, we find no evidence that AMXPs are systematically brighter compared to non-pulsing NS-LMXBs: their cumulative distributions (see Figure~\ref{fig:igr_cdf}) are almost identical.  This was previously found by \citet{Gallo2018}, who considered this over a larger X-ray luminosity range.

Perhaps the large scatter of radio luminosities of AMXPs in the investigated \lx\ range is not surprising, given their range of NS properties like spin rate, magnetic field strength, etc.. Even the presumably less complex BH-LMXBs show divergent radio behaviour in the same \lx\ range, although with smaller scatter ($\sim$0.15\,dex).

Nonetheless, we see no clear reason why \igr\ and a few other AMXPs are significantly brighter compared to the average brightness of AMXPs and non-pulsing NS-LMXBs. \citet{Tetarenko2018} investigated the influence of orbital period in ultra-compact NS-LMXBs on radio brightness, but found no correlation. Similarly, \citet{Migliari2011} found no correlation between NS spin rate and radio brightness, and \citet{Russell2018} suggested that the spin did not account for the high radio luminosity of \igr. Indeed, we find there is no clear difference in the spin rates of ``radio-luminous'' AMXPs (\igr, IGR~J17379$-$3747, XTE~J0929$-$314 and M28I) and the rest of the population.
\citet{Eijnden2018} on the other hand discovered that a highly magnetised NS binary system has significantly lower radio luminosity in comparison with the less magnetised Z-type NSs, accreting at the same X-ray luminosity. Thus, the NS magnetic field strength or its inclination with respect to the accretion disk provide an interesting avenue for further investigation \citep[e.g.,][]{Parfrey_GR_2017,Parfrey_magnsph_2017}. That said, \citet{Sanna2018} argue that \igr\ has a typical magnetic field strength for an AMXP, which would argue against the bright radio emission being driven by a low magnetic field in this system.

Another possibility is that orbital inclination may lead to different observed radio luminosities, e.g. if the system is face-on and the jet pointed towards the observer, the expected Doppler boosting may be responsible for the high radio luminosities of some AMXPs \citep[e.g.,][]{Russell2015}.  Highly beamed systems should be rare, however.  Furthermore, at least in the case of \igr, the detection of high-velocity winds \citep{Nowak2019} suggests that the system is closer to being seen edge-on \citep{Higginbottom2019}, arguing against such a scenario. We do note that the jet and disk from these systems may be misaligned \citep[e.g.,][]{2002MNRAS.336.1371M,2019Natur.569..374M}; however, this would still not adequately explain the radio emission from \igr.

Alternatively, the X-ray wind itself could be responsible for boosting radio luminosity via wind-jet interaction. This was previously suggested as a possible explanation for the observed radio flares in \sax\ \citep{Tudor2017}. Such a scenario could also be a potential explanation for the large variability of the radio emission of \igr, which appears to be the brightest during the first three radio epochs. \citet{Nowak2019} found winds using a {\it Chandra} observation that was performed 2 days before the third and brightest ATCA observation. This remains a tentative possibility, since only one high-spectral resolution observation was performed during the 2018 outburst of \igr; thus, the direct influence of X-ray winds on radio brightness could not be further explored.

Despite its large distance, \igr's radio brightness makes it a good candidate for future coordinated radio---X-ray campaigns, assuming it again goes into outburst and that the onset of the outburst is detected promptly.  Because the VLA was unfortunately in its low-resolution D-array configuration, the sensitivity of these observations was limited by significant diffuse emission in the field.  Using higher-resolution VLA, ATCA or MeerKAT observations in the future, there should be enough sensitivity to investigate (anti-)correlated radio---X-ray variations on minutes timescales, if sensitive (e.g., {\it Chandra}) and strictly simultaneous X-ray observations can be arranged.  Such observations have given important insights into the accretion regime of tMSPs in the low-luminosity accretion-disk state \citep{Bogdanov2018}, but similarly sensitive and strictly simultaneous radio---X-ray campaigns have yet to be undertaken for NS-LMXBs in full outburst --- where the phenomenology and physics is likely quite different.  At the same time, high-cadence radio---X-ray campaigns can also be valuable even if the two wavelengths are observed only quasi-simultaneously.  The radio sampling achieved in the campaign presented here missed both the pre-discovery outburst peak (MJD~$58320-58325$) as well as a later re-brightening around MJD~58380, which if better studied could have implications for understanding the radio---X-ray coupling.  If future outbursts are also multi-peaked in X-rays, then high-cadence (ideally daily) observations can also better probe whether the radio tightly tracks these X-ray variations, whether there is a lag, and whether the existence of an outflowing wind correlates with radio brightness.  

\section*{Acknowledgements}

We thank H.~Krimm for providing pre-discovery {\it Swift}-BAT measurements included here, and A.~Archibald for analysis advice. We also thank J.~Stevens and staff from the Australia Telescope National Facility (ATNF) for scheduling the DDT ATCA radio observations.  We are grateful to the staff of the GBT, and in particular T.~Minter and R.~Lynch, for scheduling DDT observations.  N.V.G. acknowledges funding from NOVA.  J.W.T.H. acknowledges funding from a Netherlands Organisation for Scientific Research (NWO) Vidi fellowship and from the European Research Council under the European Union's Seventh Framework Programme (FP/2007-2013) / ERC Starting Grant agreement nr. 337062 (``DRAGNET''). T.D.R. acknowledges support from the NWO Veni Fellowship. N.D. and J.v.d.E. are supported by an NWO/Vidi grant, awarded to N.D.. J.C.A.M.-J. is the recipient of an Australian Research Council Future Fellowship (FT140101082), funded by the Australian government. The National Radio Astronomy Observatory is a facility of the National Science Foundation operated under cooperative agreement by Associated Universities, Inc.  The Australia Telescope Compact Array is part of the Australia Telescope National Facility which is funded by the Australian Government for operation as a National Facility managed by CSIRO. This work was supported in part by NASA through the NICER mission and the Astrophysics Explorers Program. This research has made use of data and software provided by the High Energy Astrophysics Science Archive Research Center (HEASARC), which is a service of the Astrophysics Science Division at NASA/GSFC and the High Energy Astrophysics Division of the Smithsonian Astrophysical Observatory.  Lastly, we acknowledge extensive use of the NASA Abstract Database Service (ADS) and the ArXiv.

\bibliographystyle{mnras}
\bibliography{allbib}

\bsp	
\label{lastpage}
\end{document}

%% file: table1.tex
\begin{table*}
\caption[All observations]{VLA and ATCA observations \citep[including those from][i.e. epochs 1, 2 and 3]{Russell2018}, together with the corresponding
  quasi-simultaneous {\it Swift}-XRT X-ray observations and their spectral properties. All uncertainties are 1$\sigma$.  A dash is used to mark cases in which no useful constraints could be placed on the radio or X-ray spectrum.}

\begin{minipage}{180mm}
{
\renewcommand{\arraystretch}{1.5}
\begin{tabular}{@{\extracolsep{-3.2pt}}l|cccccc|cccccc@{}}
\hline\hline
\multicolumn{1}{c}{} & \multicolumn{5}{c}{Radio} & & \multicolumn{5}{c}{{\it Swift}-XRT X-ray ($1-10$ keV)} \\
\hline
\makecell[c]{Date\\ 2018} & \# & Tel. &\makecell[c]{ MJD$^b$ } & \makecell[c]{$S_{\nu}$ \\ ($\upmu$Jy)} &   \makecell[c]{$\nu$\\(GHz)} & \makecell[c]{Spectral\\ index\\$\alpha$} & &  \makecell[c]{MJD$^b$ } & \makecell[c]{Obs. ID \\ (PC mode)} & \makecell[c]{Unabsorbed$^a$ \\ flux  $\times 10^{-10}$\\($\mathrm{erg\,s^{-1}\,cm^{-2}}$)}   &  \makecell[c]{Photon \\ index} &  $\chi^2_{\nu}$ (dof)  \\
\hline
14 Aug & 1 & ATCA & 58344.56 & \makecell[c]{1090$\pm$20\\1120$\pm$20} & \makecell[c]{5.5\\9.0} &0.1$\pm$0.3& & 58344.027 & 00010804002  & 3.09$^{+0.17}_{-0.18}$ & 1.8$\pm$0.2 & 0.92 (15)\\
\arrayrulecolor{gray}\hline
19 Aug & 2 & ATCA &  58349.26 & \makecell[c]{1000$\pm$50\\850$\pm$50} & \makecell[c]{5.5\\9.0} &$-0.30\pm$0.25 & & 58348.221 & 00010804004  & 1.34$^{+0.27}_{-0.11}$ & 2.6$^{+0.3}_{-0.2}$ & 0.87 (10)\\
\arrayrulecolor{gray}\hline
25 Aug & 3 & ATCA &  58355.28 & \makecell[c]{1180$\pm$40\\1210$\pm$40\\1540$\pm$100\\1600$\pm$90} &  \makecell[c]{5.5\\9.0\\17.0\\19.0} & 0.25$\pm$0.1 & & 58355.527 & 00010804006  & 4.03$^{+0.16}_{-0.15}$ & 2.3$\pm$0.1 & 1.03 (56)\\
\arrayrulecolor{gray}\hline
1 Sep & 4 & VLA & 58362.16 & 435$\pm$144 & 10.0 & --- &  & 58362.699 & 00010804009  & 5.35$^{+0.36}_{-0.29}$ & 2.4$\pm$0.1 & 0.88 (26)\\
\arrayrulecolor{gray}\hline
28 Sep & 5 & VLA & 58389.02 & \makecell[c]{620$\pm$40\\590$\pm$20} & \makecell[c]{12.8\\17.4} &$-0.15\pm$0.23&  & 58389.058 & 00010899002  & 4.21$^{+0.20}_{-0.19}$ & 2.2$\pm$0.1 & 0.97 (32)\\
\arrayrulecolor{gray}\hline
5 Oct & 6 & ATCA & 58396.12 & \makecell[c]{480$\pm$13\\372$\pm$11} & \makecell[c]{5.5\\9.0} & $-0.50\pm$ 0.15 & & 58396.156 & 00010804015  & 1.54$^{+0.06}_{-0.07}$ & 2.1$\pm$0.1 & 1.11 (32)\\
\arrayrulecolor{gray}\hline
9 Oct & 7 & VLA & 58400.05 & \makecell[c]{568$\pm$48\\544$\pm$37} & \makecell[c]{9.0\\10.8} & $-0.24\pm$0.59 &  & 58399.941 & 00010804018  & 1.26$\pm$0.07 & 2.3$\pm$0.1 & 1.44 (26)\\
\arrayrulecolor{gray}\hline
13 Oct & 8 &  ATCA & 58404.02 & \makecell[c]{280$\pm$30\\260$\pm$25} & \makecell[c]{5.5\\9.0} &$-0.15\pm$0.29 &  &  58403.678 & 00010804020  & 0.54$^{+0.11}_{-0.08}$ & 2.6$\pm$0.5 & 1.4 (5)\\
\arrayrulecolor{gray}\hline
19 Oct & 9 & VLA & 58410.01 & <94.7 & 10.0 & --- & &\multirow{2}{*}{58410.103} & \multirow{2}{*}{00010804024} & \multirow{2}{*}{<0.02}& \multirow{2}{*}{ 2.0$^c$} & \multirow{2}{*}{---} \\
20 Oct & 10 & VLA & 58411.01 & <86.7 & 10.0 & --- & & & & & \\
\arrayrulecolor{black}\hline
\end{tabular}
}
\begin{flushleft}{
$^{a}$ We used a fixed value of $N_{\rm H} = 4.4\pm0.2 \times 10^{22}$\,cm$^{-2}$, as derived by \citet{Nowak2019}.\\
$^b$ Mid point of the observation.\\
$^c$ Fixed to determine the upper limit.
}\end{flushleft}
\end{minipage}
\label{tab:igr_obs}
\end{table*}